\documentclass[12pt,preprint]{aastex}

\begin{document}


\title{Neutrino Interactions in the Outflow from Gamma Ray Burst 
Accretion Disks}


\author{R. Surman\altaffilmark{1} and G. C. McLaughlin\altaffilmark{2}}


\altaffiltext{1}{Department of Physics and Astronomy, 
Union College, Schenectady, NY 
12308}
\altaffiltext{2}{Department of Physics, North Carolina State University, 
Raleigh, NC  27695-8202}


\begin{abstract} 

We examine the composition of matter as it flows away from gamma ray burst
accretion disks, in order to determine what sort of nucleosynthesis may
occur.  Since there is a large flux of neutrinos leaving the surface of
the disk, the electron fraction of the outflowing material will change due
to charged current neutrino interactions.  We calculate the electron
fraction in the wind using detailed neutrino fluxes from every point on
the disk and study a range of trajectories and outflow conditions for
several different accretion disk models.  We find that low electron
fractions, conducive to making $r$-process elements, only
appear in outflows from disks with high accretion rates that have a
significant region both of trapped neutrinos and antineutrinos.  Disks
with lower accretion rates that have only a 
significant region of trapped
neutrinos can have outflows with very high electron fractions,
whereas the lowest accretion rate disks with little trapping have
outflow electrons fractions of closer to one half. 

\end{abstract}




\section{Introduction}

Ever since the first gamma ray bursts were detected thirty years
ago, their origin has been a subject of great interest.  Many more
observations have occurred in recent years which point to
exotic supernova as their astrophysical source; for a review 
see \citet{mes02}.
Although the hydrodynamic details of the evolution of these objects
is still under development, see e.g. \cite{woo93,mac99}, 
it is likely that the burst
will originate from a configuration where an accretion disk surrounds
a black hole.  Several models for these accretion disks have been 
examined 
\citep{pop99,nar01,dim02}.  In addition to 
the highly collimated ultrarelativistic jet that produces the 
observed gamma rays,
there will also be outflow from the accretion disk which has been
studied by \cite{mac03,mac99}. This material
begins at relatively high temperature (a few MeV) and therefore
will undergo a primary nucleosynthesis where free nucleons form 
heavy nuclei. Since this is a different environment than that which occurs
in the relativistic jet, different nucleosynthesis products
will be produced.

Any analysis of the nucleosynthesis must begin with the
evolution of the electron fraction in the accretion disk.
This was done in \cite{pru03} by considering only electron and positron
capture and by \cite{sur04} by considering neutrino and antineutrino
capture as well. The disk is hot enough that for high accretion rates the
neutrinos and even the antineutrinos can become trapped, creating
a neutrino torus around the black hole that is in some ways similar to
the neutrinosphere at the surface of the protoneutron star in a normal
supernova.

A preliminary analysis of the nucleosynthesis from the outflow of
accretion disks was done in \cite{pru04} using a spherically symmetric
neutrino driven wind model, and by \cite{psm04} using an outflow at fixed
velocity and entropy.  \cite{maeda03} considered the nucleosynthesis that
will occur from explosive burning in gamma ray bursts. \cite{pru04}
concluded that iron peak elements will be produced in this outflow, while
\cite{psm04} examined various rare isotopes that would point to GRB disk
outflow as a unique nucleosynthesis event.  Nucleosynthesis from accretion
disks has also been discussed in \cite{fuj04}. 

In this paper we examine the nucleosynthesis in the gamma ray burst ejecta
by studying the effect of the neutrino flux on the electron fraction of
the material as it leaves the surface of the disk.  
We examine the conditions
under which the electron fraction is quite low and the heaviest,
$r$-process elements are likely to be produced.  We also examine the the
conditions under which the lighter iron peak nuclei are likely to be
formed. 

\section{Trajectories} \label{traject}

Determining the trajectories followed by mass elements ejected
from the gamma ray burst accretion disk is a complicated problem.
The trajectories depend on the initial thermodynamic parameters of the
material, and the manner in which energy is imparted to the gas.
With 
different accretion disk parameters, such as the viscosity, $\alpha$, spin 
parameter $a$ and
size of the black hole, the results are likely to vary.  Although
disk winds have been studied extensively in the context of
other astrophysical objects such as AGNs and X-ray stars, 
a complete discussion of the outflow from gamma ray burst accretion
disks has not yet been attempted.

In order to investigate the impact of neutrino interactions on 
the electron fraction and therefore  the
nucleosynthesis coming from gamma ray burst disk outflow, 
we approximate disk outflow trajectories
by using a two component approach. At very large distance,
we use spherical symmetry.   
Initially, however, the forces pushing the material off
the accretion disk will be closer to vertical and we 
assume cylindrical symmetry.
At short distances, therefore, 
the material moves approximately vertically, whereas
at long distances, it moves radially away from the black hole.  

The general hydrodynamic equations in steady state
for matter conservation, momentum conservation and energy conservation
are, e.g. \cite{dun86}:
\begin{equation}
{\partial{\rho} \over \partial{t}} = -\bigtriangledown \cdot 
(\rho {\bf u})
\end{equation}
\begin{equation}
{D {\bf u} \over D t} = - {1 \over \rho} {\bigtriangledown P -
\bigtriangledown \phi}
\end{equation}
\begin{equation}
{D \epsilon \over D t} + P {D(1 / \rho) \over D t} =
{1 \over \rho} ( \bigtriangledown \cdot {\bf S_{ph}}
+ \bigtriangledown \cdot {\bf S_\nu})
\label{eq:energy}
\end{equation}

Here $\rho$ is the density, $P$ is pressure,
 ${\bf u}$ is the velocity, $\epsilon$ is
the energy per unit mass and ${\bf S_{ph}}$
and ${\bf S_\nu}$ are the photon and neutrino fluxes respectively.
The gravitational potential is represented by $\phi$, while
the notation $D / Dt$ represents a convective derivative.

As discussed above, we divide the outflow into two regions, and
approximate the outflow in these regions as having a cylindrical or 
spherical symmetry.  For the case of spherically radial flow, we rewrite
the mass and the momentum equation, so we have:
\begin{equation}
\dot{M} = 4 \pi r^2 \rho u_r
\label{eq:sph1}
\end{equation}
\begin{equation}
u {\partial{u} \over \partial{r}} = - {1 \over \rho} 
{ \partial P \over \partial r} - {G M \over r^2}
\label{eq:sph2}
\end{equation}

In these equations $\dot{M}$ is the mass loss rate, $G$ is the
gravitational constant, and $M$ is the mass enclosed within radius
$r$. We also rewrite these equations for cylindrical geometry for vertical
flow off the disk:

\begin{equation}
\dot{M} = -2 \pi \left[ \int \rho u_r r_c dz + \int \rho u_z r_c dr_c \right]
\label{eq:cyl1}
\end{equation}
\begin{equation}
u_r {\partial{u_r} \over \partial r_c} + u_z {\partial{u_r} \over \partial{z}}
- {u_{\phi}^2 \over r_c}
= - {1 \over \rho} {\partial{P} \over 
\partial{r_c}} - G M r_c (z^2 + r_c^2)^{-3/2}
\label{eq:cyl2}
\end{equation}
\begin{equation}
u_r {\partial{u_r} \over \partial r_c}  + 
u_z {\partial{u_z} \over \partial{z}}
= - {1 \over \rho} {\partial{P} \over 
\partial{z}} - G M z (z^2 + r_c^2)^{-3/2}
\label{eq:cyl3}
\end{equation}

In the above equations $r_c$ is the radial cylindrical coordinate.

In principle one should solve Eqs. \ref{eq:cyl1} - \ref{eq:cyl3} 
(or Eqs. \ref{eq:sph1} - \ref{eq:sph2})
together with
with Eq. \ref{eq:energy}.  However, because of the complex geometry, this
involves a lengthy numerical calculation
which is not practical until more accurate disk models become
available.  Still, we would like to understand the
importance of neutrino interactions in the outflow and what sort
of electron fractions can obtain in this environment. Therefore,
in place of Eqs. \ref{eq:energy} and Eq. \ref{eq:cyl2} -\ref{eq:cyl3} 
(or Eq. \ref{eq:sph2}), we
use a parameterization for the velocity,
\begin{equation}
|u| = v_\infty \left(1 - {R_0 \over R} \right)^\beta
\label{eq:velocity}
\end{equation}
where $R = (z^2 + r_c^2)^{0.5}$ for the first, vertical part of
the trajectory and the starting position of the 
material is $R_0$.  Once we switch to spherical flow $R=r$.
We study the results as functions of the
parameters $\beta = 0.2$ to $2.5$ , 
$v_\infty =  5 \times 10^3 {\rm km} \, {\rm s}^{-1} - 
5 \times 10^4 {\rm km} \,{\rm s}^{-1}$, and $R_0 = 50 \, {\rm km}$ 
to  600 km. Although all
trajectories 
with the same $v_\infty$ asymptote to the same value at large distance, the
ones with smaller $\beta$ have a greater initial acceleration and 
arrive there more quickly.

\begin{figure}
\plotone{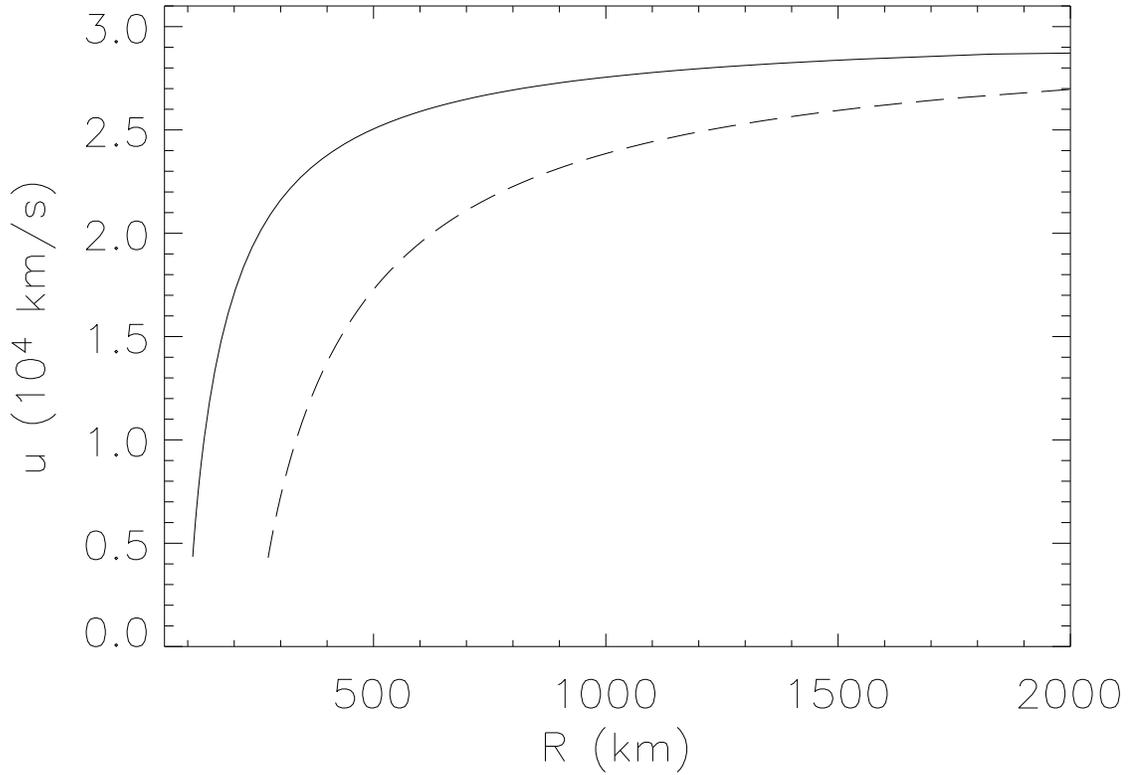}
\caption{Shows velocity plotted against 
radius for three different trajectories. The solid 
line shows Model 1 which starts at a distance on the disk from
the center of $r=100 \, {\rm km}$. The dashed line 
shows Model 2 and 3 which start at a distance of
$r=250 \, {\rm km}$ from the center.
\label{fig:velocity}}
\end{figure}

Three velocity trajectories are shown in Fig. 
\ref{fig:velocity} as a function of distance from the black hole, 
all for the same
$\beta$ and final velocity $v_\infty = 3 \times 
10^4 \, {\rm km} \, {\rm s^{-1}}$,
  but for two different initial starting points.
 
We use the mass conservation equations (Eq. \ref{eq:cyl3} or Eq.
\ref{eq:sph2}) in order to determine the density given the position and
velocity.  In the cylindrical case we assume the velocity is completely in
the $z$ direction, and therefore there is no expansion of the
material in the radial direction along the disk. 

In our calculations we assume a constant entropy, since any heating of the
material should be done at the surface of the disk.  We take the
entropy as an input parameter in the range of $s=10$ to $s=40$, 
from which we calculate the temperature
at each step using the expression for the
entropy in units of the Boltzmann constant
\begin{equation}
s = s_\gamma + s_{e^+e^-} + s_{nucleon}
\end{equation}
where $s_\gamma$ is the photon entropy, $s_{e^+e^-}$ is the entropy of 
the electron-positron pairs, and $s_{nucleon}$ is the entropy of the nucleons. 
The photon entropy is given by
\begin{equation}
s_\gamma = 0.019 {T_{MeV}^3 \over \rho_{10}}
\end{equation}
where $T_{MeV}$ is the temperature in units of MeV and $\rho_{10}$ is
the density in units of $10^{10} \, {\rm g} \, {\rm cm}^{-3}$.  The 
%and the 
entropy of the electron positron pairs is
\begin{equation}
s_{e^+e^-} = 0.0022 {T_{MeV}^3 \over \rho_{10}} \left[ F \left({\mu_e \over T_{MeV}}\right) + F\left(-{\mu_e \over T_{MeV}}\right)
\right]
\end{equation}
\begin{equation}
F\left({\mu_e \over T_{MeV}} \right) =
\int_0^\infty dx
{x^4 /(3y)  + x^2 y - (\mu_e / T_{MeV})  x^2 
\over 1 + \exp(y - {\mu_e / T_{MeV}})}
\end{equation}
where $y = (x^2 + (m_e/T_{MeV})^2)^{0.5}$, $m_e$ is the mass of the electron
and 
$\mu_e$ is the chemical potential of the electrons, which is determined
by the temperature, density and electron fraction of the material
\begin{equation}
\rho_{10} Y_e = 2.2 \times 10^{-3} 
\left[F_{2m} \left({\mu_e \over  T_{MeV}} \right) - F_{2m}\left(-{ \mu_e \over T_{MeV}}\right) \right]
\end{equation}
where
\begin{equation}
F_{2m}\left({\mu_e \over T_{MeV}}\right) = \int^\infty_0 {x^2 dx \over 1 + \exp(y - \mu_e / T_{MeV}) }.
\end{equation}
We take the complete expression for the electron and positron
pairs since we are neither in the limit where they are fully relativistic or
fully non-relativistic.  The approximate expression for the entropy of the
nucleons is
%uses Y_e = 0.5
\begin{equation}
s_{nucleon} = 7.4 + \ln \left({T_{MeV}^{3/2} \over \rho_{10}} \right).
\end{equation}

We begin the calculations when the material is at the surface of the disk.  
We start with disk conditions from the disk models of \citet{dim02} 
(hereafter DPN) for 
disks with accretion rates $\dot{M} \ge 1 \, {\rm M}_{\sun} / s$ and 
from \cite{pop99} for more slowly accreting disks. 
We take the disk surface to be at the density scale height 
\begin{equation}
H =  |{1\over \rho} {d \rho \over dz}  |^{-1}.
\end{equation}
We begin the outflow in the
vertical direction so $u_r$ is zero, and $u_z = |u|$.  We take steps
in vertical distance, at each point determining a velocity, a density,
temperature and electron chemical potential.  We turn from the vertical
solution to the cylindrical one when the material has reached one, 
two or three  vertical
scale heights above the disk, although this is the least sensitive parameter
in this model.

\begin{figure}
\plotone{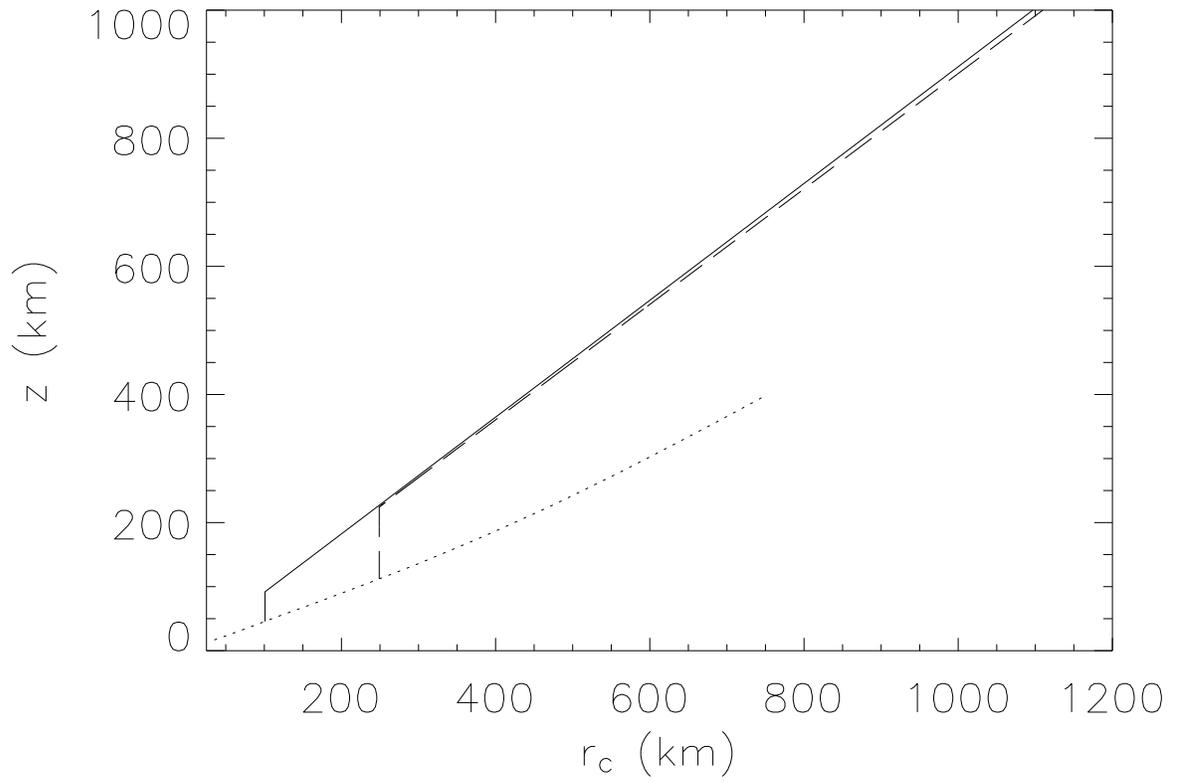}
\caption{Shows vertical height 
as a function of radial coordinate, $r_c$, for the
same three models as in Fig. 
\ref{fig:velocity}.  The dotted line shows the scale
height of the disk. 
\label{fig:xz}}
\end{figure}

\begin{figure}
\plotone{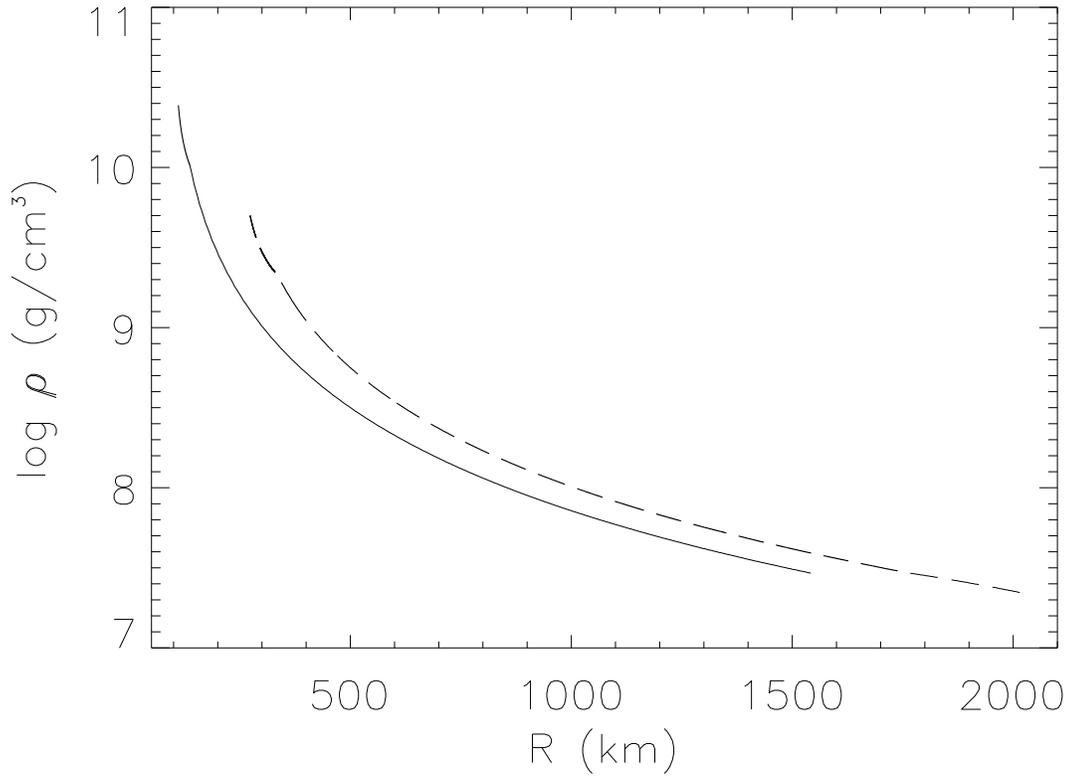}
\caption{Shows density as a function of distance from the center,
R = $(z^2 + r_c^2)^{0.5}$, for the same three models 
as in Fig \ref{fig:velocity}. \label{fig:rho}}.
\end{figure}

\begin{figure}
\plotone{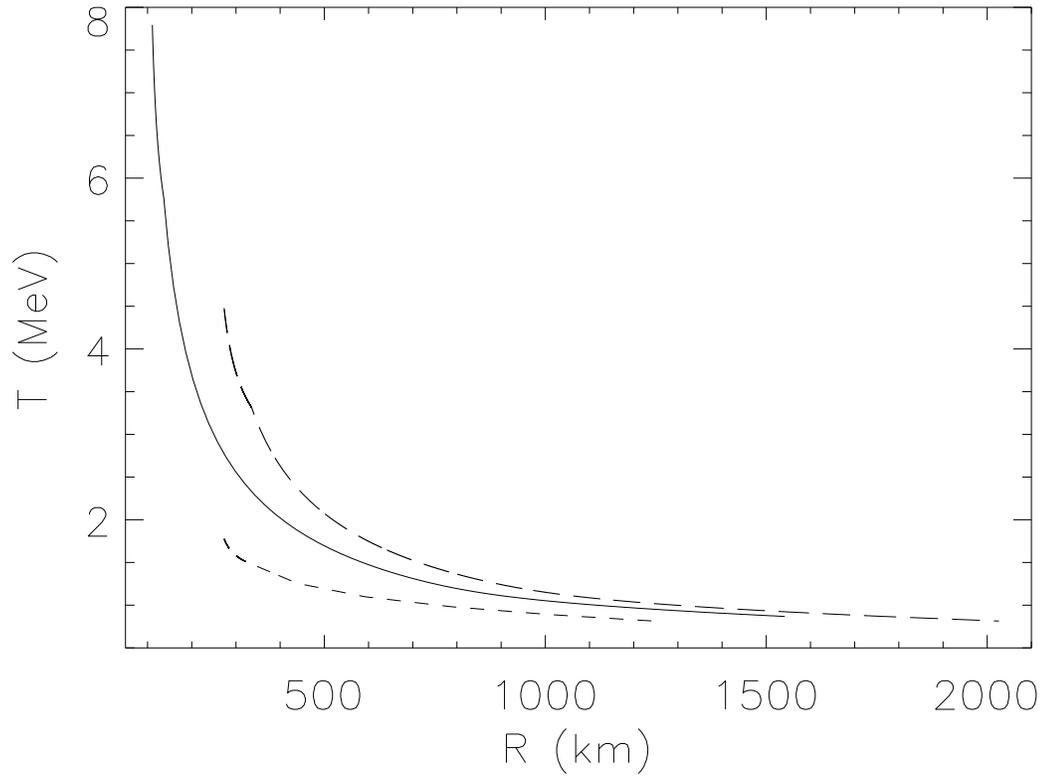}
\caption{Shows temperature as a function of R, 
for Model 1 (solid line), Model 2 (short dashed line) and
Model 3 (long dashed line).  \label{fig:t}}.
\end{figure}

The results of this formulation for a relatively high accretion rate disk model
and wind parameters,
$\beta= 0.8$ and $r_c = 100 \, {\rm km}, 
s= 20$ (Model 1), $r_c = 250 \, {\rm km},
s= 10$ (Model 2) and  $r_c = 250 \, {\rm km},
s= 20$ (Model 3) are shown in Figs. \ref{fig:xz},
\ref{fig:rho} and \ref{fig:t}.  
In Fig. \ref{fig:xz}, we plot two sample trajectories in position space.
Since we have fixed the turnover point at two vertical scale heights, the
final trajectories appear to lie almost on top of each other.  As  mentioned
above, we found little impact on the results when examining different turnover
points.  In Fig. \ref{fig:rho}, we show the densities for these 
trajectories. 
%Trajectories which begin closer to the center of the disk will 
%have higher entropy and temperature (Fig. \ref{fig:t}).  
In
Fig. \ref{fig:t} the effect of a larger entropy can also be seen as a 
greater temperature at a given distance.

\section{Neutrino Fluxes} \label{fluxes}

Next we use the trajectories developed in the previous section to 
calculate the evolution of the electron fraction due to electron,
position, neutrino and antineutrino annihilation:
\begin{equation}
e^- + p \leftrightarrow \nu_e + n
\label{eq:ecap}
\end{equation}
\begin{equation}
e^+ + n \leftrightarrow \bar{\nu}_e + p
\label{eq:poscap}
\end{equation}
Since the electrons and positrons are in equilibrium with the baryons
and photons and therefore their distribution can be described with
the temperature, $T$, and electron chemical potential $\mu_e$, the
forward rates in  Eqs.\ref{eq:ecap} and \ref{eq:poscap}
are easy to calculate.  However,
the neutrinos are not in equilibrium and their flux at each point on
the trajectory must be calculated by summing over the neutrinos
which originate at every part of the disk.

In  \cite{sur04},
we calculated the evolution of the electron fraction of a mass element
as it spiraled from the outer edge of the toward the center. This was done
by summing all contributions form the neutrino flux at all points on the disk,
taking into account regions where the neutrinos become optically thick.
These calculated electron fractions are the starting $Y_e$s for
the material ejected from the disk.  Furthermore, these neutrino fluxes
are used to calculate the spectrum at every point above the disk through
which the ejected material passes.

\begin{figure}
\plotone{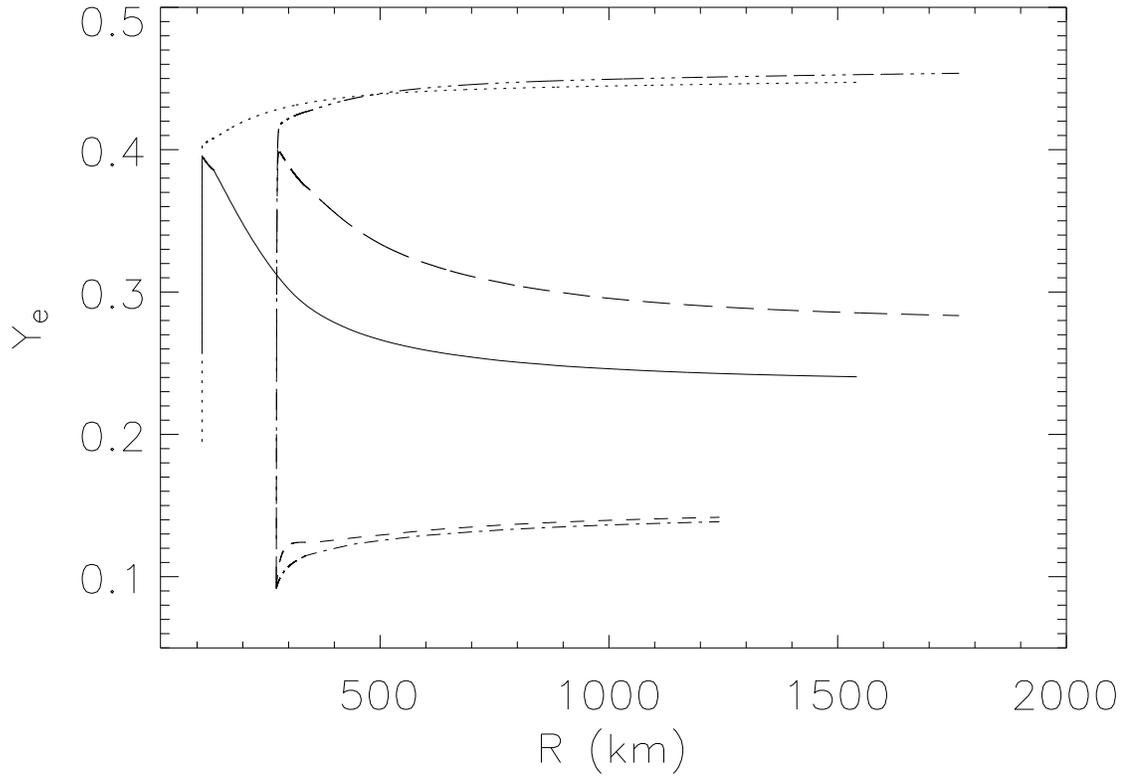}
\caption{Shows $Y_e$ as a function of distance from the black hole for the
same three models as in Fig. \ref{fig:velocity}. The dotted and dot-dashed 
lines show the
effect when neutrino capture interactions are turned off.
\label{fig:Ye}}
\end{figure}

In Fig. \ref{fig:Ye} we show the results of using the reactions
Eqs. \ref{eq:ecap} and \ref{eq:poscap} to determine the electron
fraction for Models 1, 2, and 3.
Dashed lines 
show the true electron fraction while the dot-dashed lines show the electron
fraction calculated without neutrino capture interactions.  

In the high entropy models ($s=20$) 
the large electron fractions are due to
the increased importance of electron and positron capture at higher
temperature.  With higher temperatures the electrons and positrons
essentially reset the electron fraction to a new equilibrium value and
the material ``forgets'' its original disk value and winds up
consisting of nearly equal numbers of neutrons and protons.  
This is evident in
the initial sharp increase in the electron fraction.

However, when the neutrino interactions have been included, the situation is
quite different.  Even at these high entropies, the neutrinos have a marked
impact on the evolution of the electron fraction, bringing it down to  
$Y_e < 0.3$.  The reverse rates in Eq. \ref{eq:ecap} and \ref{eq:poscap} 
overwhelm the forward rates.  Furthermore, due to the higher energies of
the electron antineutrinos, the antineutrino capture on protons is larger
than neutrino capture on neutrons.  As can be seen from the figure,
the release point of the material determines the degree to which
neutrino capture influences the final electron fractions.

For lower entropies we find less of an increase in the electron fraction
when
 the neutrino capture reactions are turned off.  The lower 
entropy means that the system is partially electron degenerate, and so even 
while the electron and positron captures are maintaining a weak equilibrium
by themselves, the electron fraction is quite low.  Note that the 
low entropy model corresponds to essentially no heating of the outflow
material since the disk has an entropy of order 10.

\section{Results}

In this section we explore the qualitative effect of variables such
as entropy and outflow timescale  on the electron fraction and 
therefore on the nucleosynthesis.  We also explore different
disk models.

\begin{figure}
\plotone{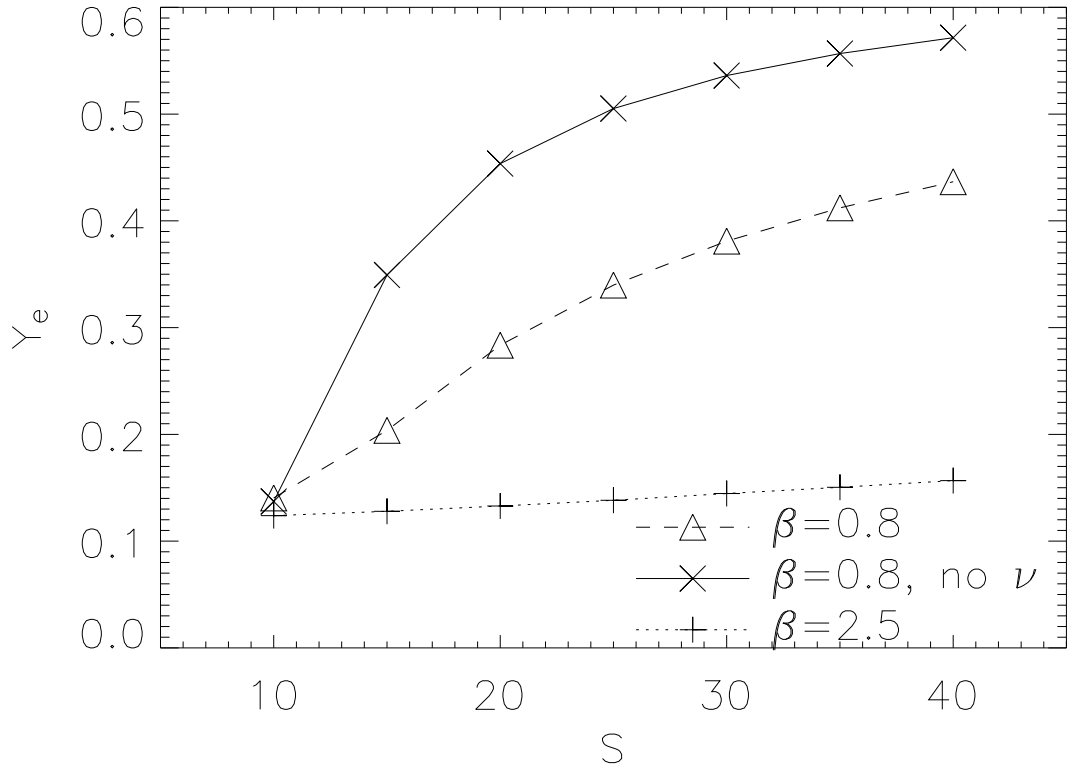}
\caption{For disk models with  $\dot{M} = 10\, {\rm M}_\sun \, \ {\rm s}^{-1}$, 
$a=0$.  Shows $Y_e$ as a function of $s$, using a
release point of $r_0 = 250 \, {\rm km}$ and a final velocity
of $v_\infty = 3 \times 10^4 {\rm km} \, {\rm s}^{-1}$.
The neutrinos have the maximum effect in a slowly accelerating 
outflow ($\beta = 2.5$) and
significantly decrease
the electron fraction.
\label{fig:yevss}}
\end{figure}

In the previous section we discussed the importance of entropy
on the electron fraction.  The outflowing material becomes less
neutron rich if it is exposed to more positron capture, which happens
when the entropy rises and the chemical potential decreases.
Fig. \ref{fig:yevss} shows the electron fraction measured at 
${\rm T}_9 = 10$
for several different entropies, for trajectories that start at 
$r_0=250 \, {\rm km}$ and two different outflow
parameters $\beta$.  This figure shows that at the highest entropies
$s=40$ and fast accelerations $(\beta = 0.8)$, 
where the neutrinos have the least influence, 
the electron fraction can become as high
as 
%(Y_e only gets as high as 0.6 when the neutrinos are neglected)
0.45, while for slow accelerations ($\beta = 2.5$) 
the neutrinos insure that the electron
fraction is very low ($\sim 0.1$).  

\begin{figure}
\plotone{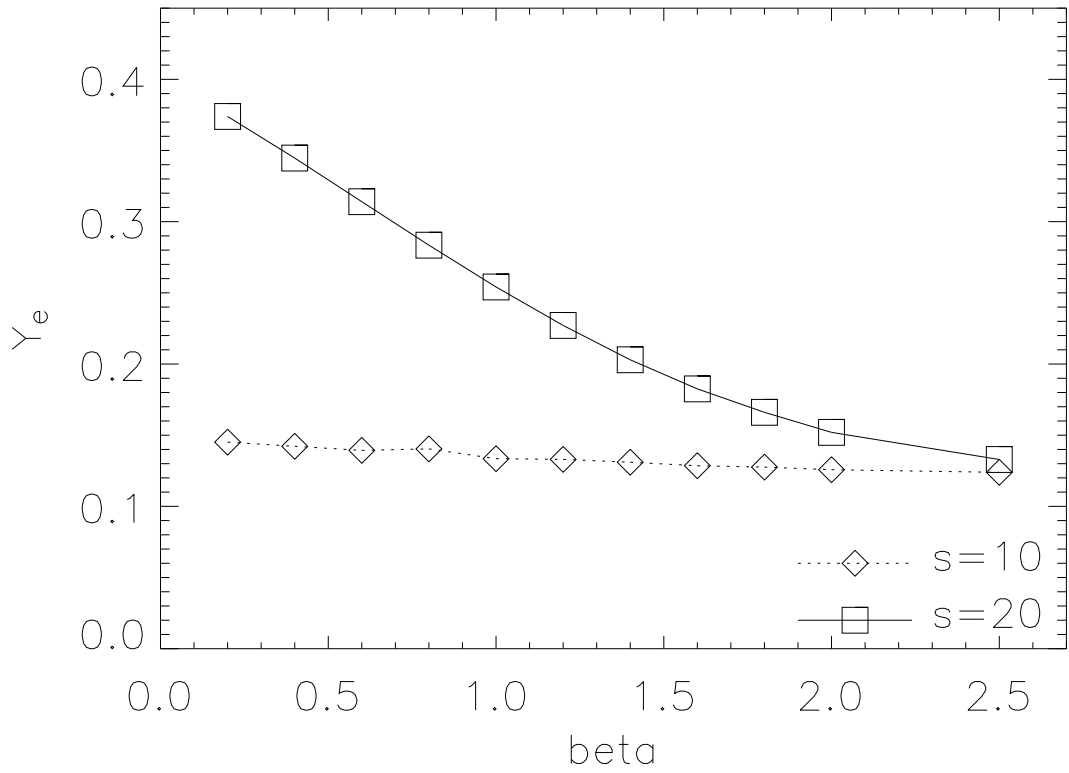}
\caption{For disk models with $\dot{M} = 10 \, {\rm M}_\sun \, \ {\rm s}^{-1}$, $a=0$.
Shows $Y_e$ as a function of $\beta$
at a release point of $r_0 = 250 \, {\rm km}$ and a final velocity
of $v_\infty = 3 \times 10^4 {\rm km} \, {\rm s}^{-1}$.  
\label{fig:yevsbeta}}
\end{figure}

The effect of the time scale of the outflow through the
parameter $\beta$ is shown in Fig. \ref{fig:yevsbeta}.  The material
accelerates much more quickly at lower $\beta$ providing less opportunity
for the neutrinos to move the system toward weak equilibrium.  In the case of
the high entropy, the effect is most pronounced.  The higher betas mean
more time for the neutrinos to drive the electron fraction down.  In the
case of the lower entropy, as previously noted the electrons and positrons
already favor a low electron fraction, and so the electron fraction 
never gets very high.

So far we have considered disk models with accretion rates of 
$\dot{M} = 10 \, {\rm M}_{\sun} / s$,
and spin parameter $a=0$. This 
model is desirable because it produces 
many neutrinos which create a large energy deposition from 
neutrino-antineutrino annihilation.  Such high accretion
 rates may be expected in the case of neutron star-neutron star mergers,
but lower accretion rates are suggested for the collapsar model
\citep{mac99}.  However,  
increasing the spin parameter or viscosity is similar to increasing the 
accretion rate as far as driving up the neutrino flux. So the high accretion
rate model we have 
considered may mimic a lower accretion rate model with a large spin parameter.

\begin{figure}
\plotone{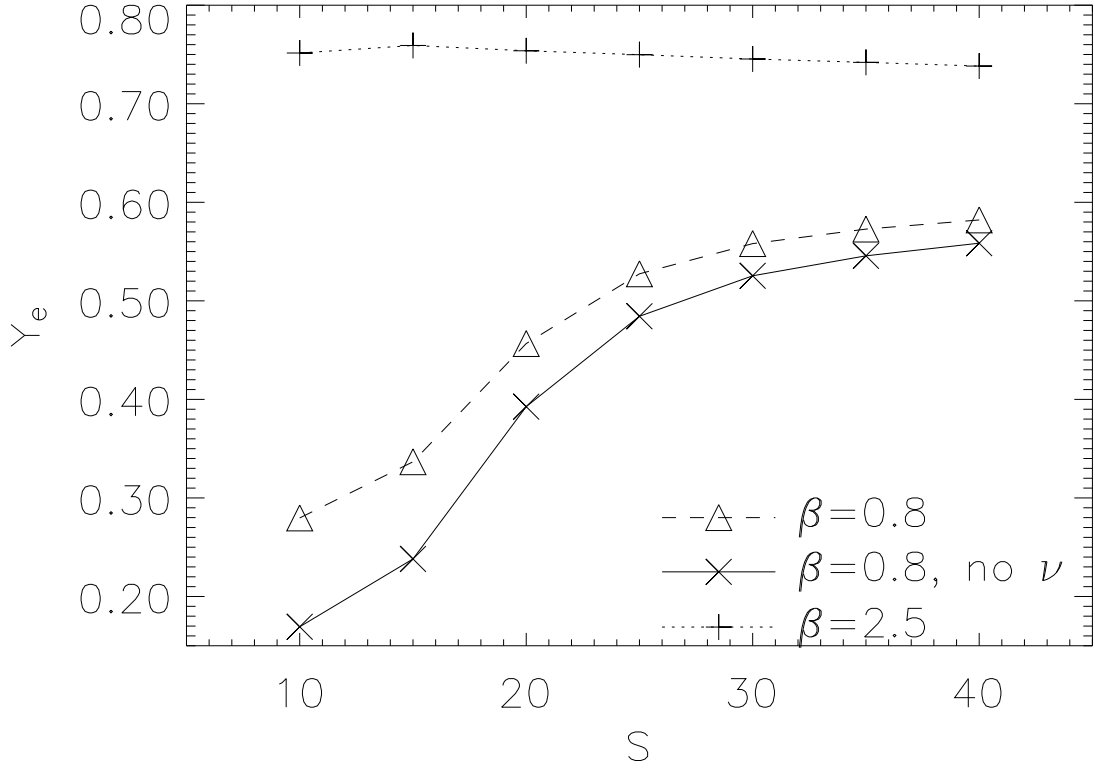}
\caption{For a disk model with an accretion rate of 
$\dot{M} = 1.0 \, {\rm M}_\sun \, \ {\rm s}^{-1}$, $a=0$.
Shows $Y_e$ as a function of $s$, using a
release point of $r_0 = 170 \, {\rm km}$ and a final velocity
$v_\infty = 3 \times 10^4 {\rm km} \, {\rm s}^{-1}$.  The neutrinos
raise the electron fraction considerably in a slowly accelerating outflow 
($\beta = 2.5$).    
\label{fig:mdot1}}
\end{figure}

When the accretion rate becomes lower an interesting effect occurs.
At $\dot{M} = 10 \, {\rm M}_{\sun} / \, {\rm s}$, $a=0$ the neutrino 
surface is at 200 km
but by $\dot{M} = 1\, {\rm M}_{\sun} / \, {\rm s}$, $a=0$, it has shrunk 
to 40 km.  Similarly the
antineutrino sphere shrinks from 140 km to around 32 km; see figures 3 and 4
in \cite{sur04}.  There are relatively few antineutrinos because the 
antineutrino surface is quite small,
however the neutrino capture rates are still large
and the net effect can be to drive $Y_e$ to very high values.
%, e.g $Y_e \sim 0.6$.
This can be seen in Fig. \ref{fig:mdot1} which 
gives the electron fraction for a $\dot{M} = 1 \, {\rm M}_{\sun}/ \, {\rm s}$ 
model for various entropies both with neutrinos and without neutrinos.
Note the difference in the slow acceleration ($\beta = 2.5$) curves in
Figs. \ref{fig:yevss} and \ref{fig:mdot1}.

This model is most sensitive to the outflow parameters because the
neutrino capture rates, while still quite strong, are smaller
than in the $\dot{M} = 10 \, {\rm M}_{\sun} / \, {\rm s}$ model, 
by a factor of 2 - 3
 for the neutrinos and an order of magnitude for the antineutrinos. 
Because of the extreme sensitively to the neutrino
parameters a neutrino diffusion calculation is needed to better determine the 
spectra and luminosity of the neutrinos and antineutrinos emitted at
each part of the accretion disk. 

\begin{figure}
\plotone{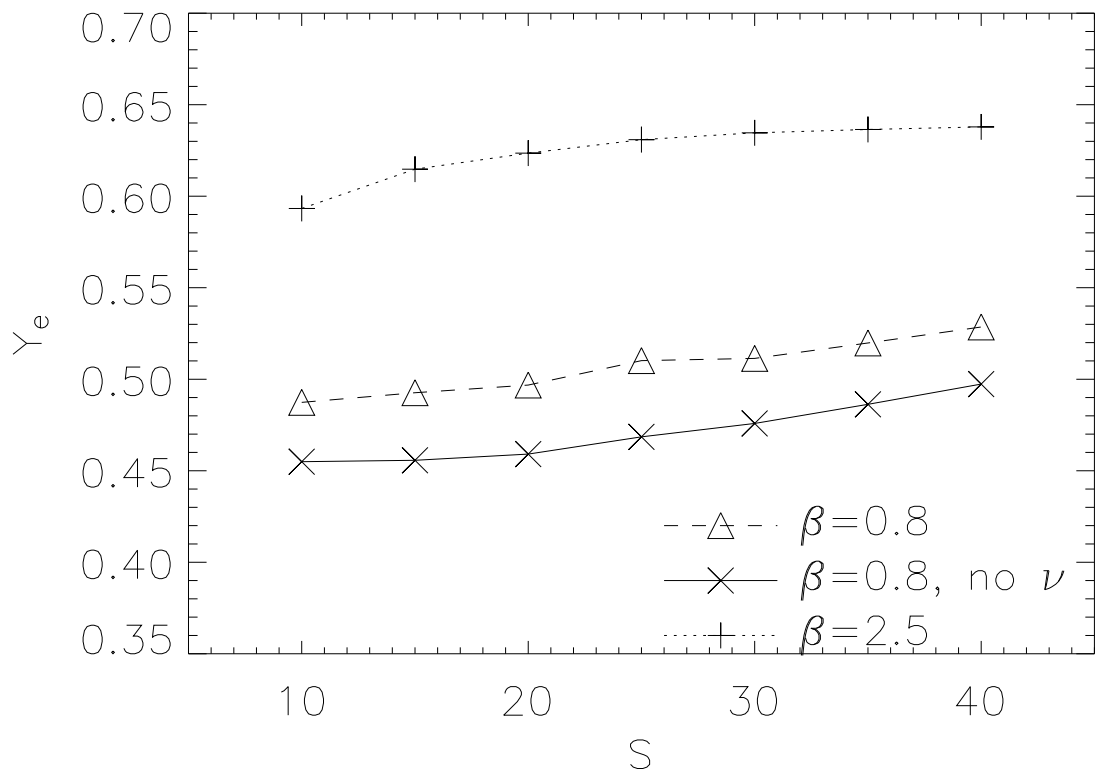}
\caption{For a disk model with an accretion rate of 
$\dot{M} = 0.1 \, {\rm M}_\sun \, \ {\rm s}^{-1} $,
spin parameter $a= 0.95$.
Shows $Y_e$ as a function of $s$, using a
release point of $r_0 = 100 \, {\rm km}$ and a final velocity
$v_\infty = 3 \times 10^4 {\rm km} \, {\rm s}^{-1}$. The neutrinos
raise the electron fraction.  For slowly accelerating outflows 
($\beta = 2.5$), the effect can be as large as 30\%.
\label{fig:mdot.1}}
\end{figure}

In Fig. \ref{fig:mdot.1}, electron fractions for models with lower
accretion rate disks of $\dot{M} = 0.1 \, {\rm M}_{\sun} / \, {\rm s}$, $a=0.95$ 
are shown.  In this
model the material starts off with a low density and temperature relative to
that of the higher accretion rate models.  
Although
 there is only a very small region where the neutrinos are trapped 
in this disk, and the antineutrinos are not trapped at all, the influence of 
neutrino
capture is apparent.  Even here, the electron
fraction changes by as much as 30\% in the upward direction depending on the
model.  Such changes will have an important impact on the nucleosynthesis
in the iron peak region.
  
Since the parameterization discussed here is independent of
the heating mechanism,
these results 
can be used as a rough gauge of the electron fraction and therefore the
nucleosynthesis for any type of wind model.

\subsection{Preliminary Nucleosynthesis Calculation}

\begin{figure}
\plotone{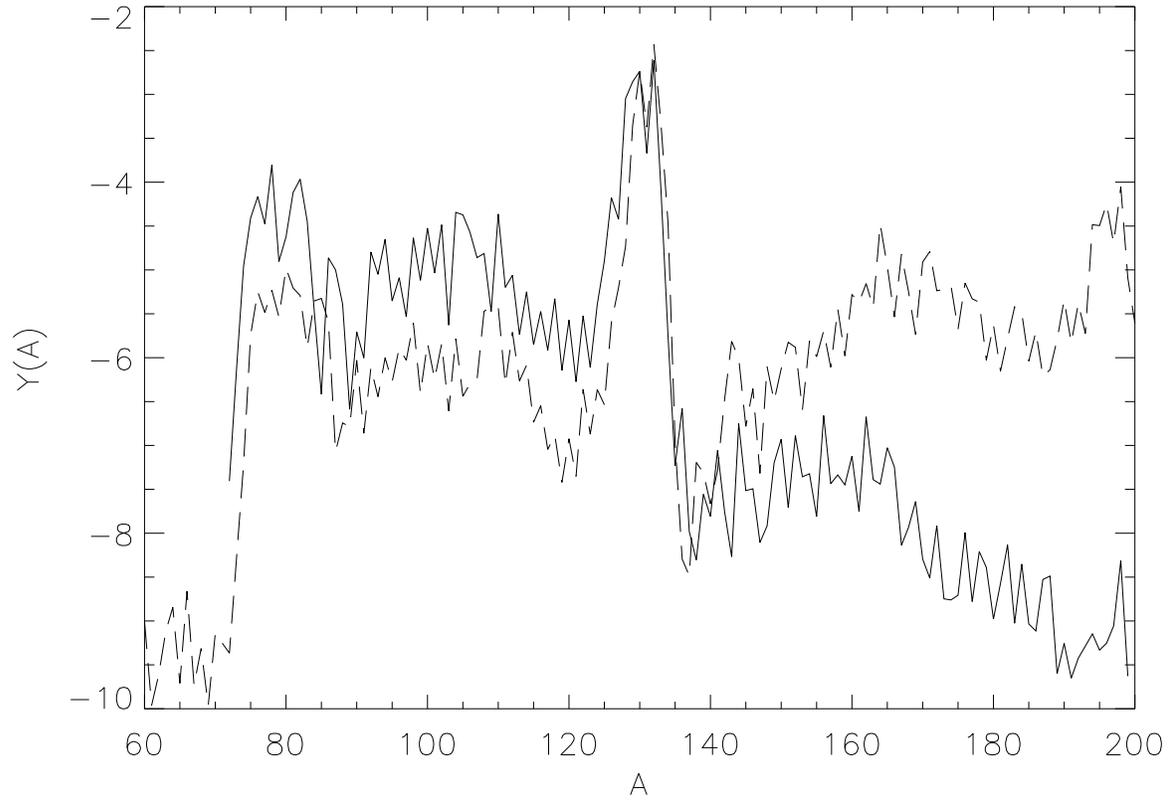}
\caption{For a disk model with an accretion rate of $\dot{M} = 10
\, {\rm M}_\sun \, \ {\rm s}^{-1} $,
spin parameter $a= 0$ and for a trajectory of $s=10$, $\beta = 0.8$,
$v_\infty = 1 \times 10^4 {\rm km} \, {\rm s}^{-1}$ and 
$r_0 = 250 \, {\rm km}$.
The solid line gives the result with all neutrino interactions in the
wind turned on, the dashed line has neutrino interactions in the disk,
but not in the wind.
\label{fig:rprocess}}
\end{figure}

The nucleosynthesis that may be produced from configurations
that give a low electron fraction is shown in figure \ref{fig:rprocess}. 
Here we show the results of an 
$r$-process calculation for $s=10$, $\beta=0.8$ and $r_0= 250 \, {\rm km}$.  
For this choice of parameters, the outflow does result in $r$-process 
abundances.  However, it can be seen from the figure that although the 
neutrinos initially helped to keep the electron fraction low, they
actually cause the A=195 peak to disappear in an alpha effect \citep{fm95,mfw96}.

The neutrino driven wind models for low accretion rate disks ($\dot{M} = 0.1 M_\sun \, / {\rm s})$ were
discussed in \cite{pru04}. These 
fall at around $\beta = 2.5$, 
$v_\infty = 3 \times 10^4 {\rm km} \, {\rm s}^{-1}$ in our 
parameterization and we estimate that the electron fraction
may be increased by as much as 5\% - 20\% by the neutrinos.  These conditions
may lead to 
a large overproduction of  elements such as $^{42}{\rm Ca}$ and
$^{45}{\rm Sc}$, $^{46}{\rm Ti}$, $^{49}{\rm Ti}$, $^{63}{\rm Cu}$,
$^{64}{\rm Zn}$ as discussed in \cite{psm04}.

For winds that accelerate quickly from an accretion disk of around 
$\dot{M} = 1 {\rm M}_\sun/s$
% = 1$
 and $a=0$, the electron fraction can become
as high as 0.8.  In this case the nucleosynthesis would be dominated by nickel,
in addition to making nuclei on the proton rich side of the valley of beta
stability, such as $^{58}{\rm Cu}$, $^{59} {\rm Zn}$,$^{50}{\rm Fe}$
and $^{52}{\rm Fe}$.

\section{Conclusions}

We have presented a parameter study for outflow from gamma ray burst accretion
disk and calculated the electron fractions produced in these outflows. 
We have considered
the impact of charge changing neutrino interactions on the outflow,
using our previously calculated neutrino fluxes from every point on the disk
\citep{sur04}.

Complete hydrodynamic models for the outflow for various disk
parameters will become available in the future and the parameter
study presented here can then be used to determine which disk models
and outflows
are likely to produce different types of nucleosynthesis, such
as $r$-process or iron peak nuclei.  For example, we find that the conditions
that are most conducive to making the $r$-process elements come from 
those disks with
high accretion rates or spin parameters, e.g. $\dot{M} = 10 {\rm M}_\sun /s$,
$a=0$, 
such that they produce a
sizable region of trapped antineutrinos.  Disk models with somewhat 
lower accretion
rates and spin parameters, e. g. $\dot{M} = 1 {\rm M}_\sun /s$,
$a=0$ may still have a significant region of 
trapped neutrinos, but a smaller region of trapped antineutrinos. This
causes the electron fractions to become very high, potentially as high as 0.8.
Still lower accretion rate models, e.g. $\dot{M} = 0.1 {\rm M}_\sun /s$,
$a=0.95$ will have have very small region of trapped neutrinos, which can
raise the electron fraction by 5\% to 30\%.

Several parameters determine the electron fraction in
the outflow, the outflow timescale,
the entropy and the release point on the disk.
The effect of the high entropy
is to raise the electron fraction through positron capture.
A more slowly accelerating wind  
increases the importance of neutrino and antineutrino capture which can drive
the electron fraction up or down. Releasing the material closer to the center
of the disk has a similar effect.

In all cases there is interesting nucleosynthesis to explore.  However,
if the $r$-process is to come from gamma ray bursts, it is necessary 
for some of the outflow material to have a small electron fraction.
This requires high accretion rate, high spin models, and small amounts
of heating in the wind.

Future observations of emission lines from gamma ray bursts, when
combined with studies such as this one may be an avenue toward
understanding not only the nucleosynthesis originating from GRBs but also the
conditions in the outflow and in the accretion disk itself.

\acknowledgments

We wish to thank A. ud-Doula and J. Pruet for useful discussions.  
G. C. M. acknowledges 
support from the US Department of Energy under grant DE-FG02-02ER41216, and 
R. S. acknowledges support from the Research Corporation under grant CC5994.

\clearpage


\begin{deluxetable}{ccccccccc} 
\tablewidth{0pt}
\tablecaption{Disk and Wind Parameters for Three Sample Trajectories}
\tablehead{
\colhead{Model} & Disk Model & \colhead{$\dot{M}$} & \colhead{$a$} & \colhead{$\alpha$} 
& \colhead{s} & \colhead{$\beta$} & $R_0$
& \colhead{$v_\infty$} 
}
\startdata
1 & DPN & $10 \, {\rm M}_{\sun} / {\rm s}$& 0 & 0.1 & 20 & 0.8 &  100 km  &  $3 \times 10^4$ km /s \\
2 & DPN & $10 \, {\rm M}_{\sun} / {\rm s}$& 0 & 0.1 & 10 & 0.8 &  250 km  & $ 3 \times 10^4$ km /s \\
3 & DPN & $10 \, {\rm M}_{\sun} / {\rm s}$& 0 & 0.1 & 20 & 0.8 &  250 km  & $ 3 \times 10^4$ km /s \\

\enddata
\end{deluxetable}

%% If you are not including electonic art with your submission, you may
%% mark up your captions using the \figcaption command. See the 
%% No more than seven \figcaption commands are allowed per page, 
%% so if you have more than seven captions, insert a \clearpage 

%% The following command ends your manuscript. LaTeX will ignore any text

\end{document}